# Deep Learning-Based 3D Seismic Velocity Inversion Under Dual-Domain Sparse Representation

Guoxin Chen[1], Wenjie Wang[1], Haiyang Lu[1], Jinxin Chen[1]

1 Ocean College, Zhejiang University, Zhoushan, China, 316021

**Abstract**: Three-dimensional seismic full-waveform inversion (FWI) provides high-fidelity subsurface velocity models but is restricted by high computational cost, strong nonlinearity, cycle-skipping, and heavy dependence on initial models. Although data-driven deep learning mitigates these issues, it still produces over-smoothed results with limited physical interpretability and low efficiency. To address these challenges, we propose a dual-domain sparse deep learning framework for 3D seismic velocity inversion using the discrete cosine transform (DCT). DCT compresses seismic data and velocity models into a sparse domain to remove redundancy while preserving key structural features. A geometry-adaptive network named SEDCN (Squeeze-and-Excitation Deformable Convolutional Network) is adopted to better capture irregular salt-dome geometries and sharp velocity boundaries. We train and validate the network on 676 samples from the 3D SEG/EAGE salt model, with two schemes for comparison: the proposed DCT-SEDCN and the baseline SEDCN without DCT. Numerical results show that DCT-SEDCN reduces training time by more than 90% and achieves higher PSNR and SSIM than conventional spatiotemporal-domain methods. It effectively suppresses over-smoothing, recovers salt body boundaries and stratigraphic details clearly, and generates geologically more reliable velocity models. This study confirms that DCT-based sparse representation combined with geometry-adaptive deep learning significantly improves the efficiency, accuracy, and robustness of 3D seismic velocity inversion. The framework offers a scalable solution for large-scale 3D FWI and can be extended to elastic/viscoelastic multi-parameter inversion and field data applications.

## Introduction

The primary research directions of 3D seismic full-waveform inversion (FWI) focus on improving computational efficiency, enhancing inversion convergence, and developing practical application strategies for field data. Computational efficiency is mainly improved through hybrid-domain strategies, such as the time-frequency domain hybrid scheme (Tran et al., 2020) and the time-Laplace-Fourier hybrid approach (Kim et al., 2013). Additional improvements include hardware acceleration (Liu et al., 2015), wavefield reconstruction and data compression (Raknes et al., 2016), and grid optimization strategies (Lee et al., 2013; Wang et al., 2019; Plessix, 2009). Convergence enhancement is achieved via objective function refinement (Métivier et al., 2016; Borisov et al., 2018), uncertainty quantification (Zhang et al.,

2023), refined physical modeling (Yang et al., 2017; Mirzanejad et al., 2019; Trinh et al., 2019), and multi-parameter cooperative inversion (Marjanović et al., 2019; Yao et al., 2025). These methodological advances have promoted the successful application of model-driven inversion in near-surface engineering (Irnaka et al., 2022), oil and gas reservoirs and complex structures (Boddupalli et al., 2021), crustal structure and mid-ocean ridge exploration (Morgan et al., 2015; Fichtner et al., 2013), as well as time-lapse monitoring and survey optimization (Abdellaziz et al., 2025; Mercier et al., 2025).

Although remarkable progress has been achieved in model-driven inversion, challenges still exist in handling complex structures, strong nonlinearity, the lack of low-frequency components, and strong dependence on initial models. These issues have motivated researchers to explore data-driven methods such as deep learning. At present, deep learning has been intensively investigated and widely applied in geophysical exploration, with remarkable achievements in seismic data preprocessing (Ovcharenko 2021; Chen et al., 2024; Qi et al., 2025, 2026; Du et al., 2024) and full-waveform inversion (Virieux & Operto 2009; Ten Kroode et al. 2013; Chen et al., 2017, 2018a, b, 2025; Yang & Ma 2019; Rasht-Behesht et al. 2022). In 3D seismic inversion, current deep learning studies mainly focus on addressing the challenges of computational efficiency, cycle-skipping, and complex modeling. Bastidas Pérez (2022) combined a genetic algorithm (GA) with an entropy-regularized FWI strategy and used deep learning to refine salt body models; Zeng et al. (2021) designed the memory-efficient InversionNet3D for high-resolution reconstruction; Celaya and Araya-Polo (2023) developed a multimodal joint inversion framework to improve monitoring accuracy; Ren et al. (2021) constructed an automated model generation framework to provide synthetic datasets; Gelboim et al. (2023) introduced compressive learning for efficient inversion; Zhang et al. (2022) and Zhu et al. (2023) designed networks to reconstruct density distributions and perform joint gravity inversion to reduce errors, respectively. Studies demonstrate that deep learning architectures embedded with physical constraints can significantly improve the efficiency and robustness of 3D seismic inversion. Nevertheless, data-driven inversion approaches still suffer from several critical limitations, including poor interpretability and insufficient physical consistency of the models, heavy reliance on large volumes of high-quality training data, limited generalization ability, and extremely high computational resource consumption during training.

To overcome these limitations, we propose a feature-domain deep learning framework that combines Discrete Cosine Transform (DCT) compression with a geometry-aware network architecture. DCT leverages the sparsity of seismic signals to compress the raw data, eliminating redundant features while preserving important geological information (Chen et al., 2024; Chen et al., 2025; Rincón et al., 2025). Crucially, this compression shifts the network' s focus from noisy temporal samples to structured frequency representations. Furthermore, we design a deformable ResNet backbone network and incorporate a squeeze-excitation (SE) attention mechanism—deformable convolutions adapt the sampling grid to irregular salt body geometries, while SE modules dynamically amplify structurally critical frequencies. Validated on the SEG/EAGE 3D salt model demonstrating unprecedented efficiency and fidelity in effectively recovering complex salt body structures. This work shows that the combination of physically guided feature compression and adaptive networks is crucial for scalable 3D seismic inversion.

**Methodology**

As shown in Figure 1, this study proposes a dedicated feature extraction network named SEDCN (Squeeze-and-Excitation Deformable Convolutional Network) for 3D seismic velocity inversion. It is built on a fully convolutional network (FCN) as the base architecture, which preserves complete spatial structural information and supports end-to-end prediction without fully connected layers. The SEDCN is an independent geometric feature learning module that has no direct connection with the DCT sparse domain transformation, and its core role is to enhance the representation ability of complex geological structures.

The SEDCN is constructed as a deformable residual network, which integrates Deformable Convolutional Networks (DCN) and Squeeze-and-Excitation (SE) attention as key components. Deformable convolution enables the network to dynamically adjust its receptive field and sampling points according to the geometry of salt domes, layer boundaries, and velocity interfaces, thereby significantly improving adaptability to irregular structures and sharp velocity variations. The SE attention mechanism performs channel-wise feature recalibration to automatically enhance effective structural features while suppressing noise and redundant information. Residual connections are embedded in each module to avoid gradient vanishing during deep network training, stabilize the training process, and accelerate convergence.Benefiting from the synergistic effect of deformable convolution, channel attention, and residual learning, the SEDCN exhibits stronger structural perception, higher feature representation capability, and better robustness for complex salt dome models. This design ensures that the network can accurately capture subtle structural details and boundary information, laying a solid foundation for high-precision 3D seismic velocity reconstruction.

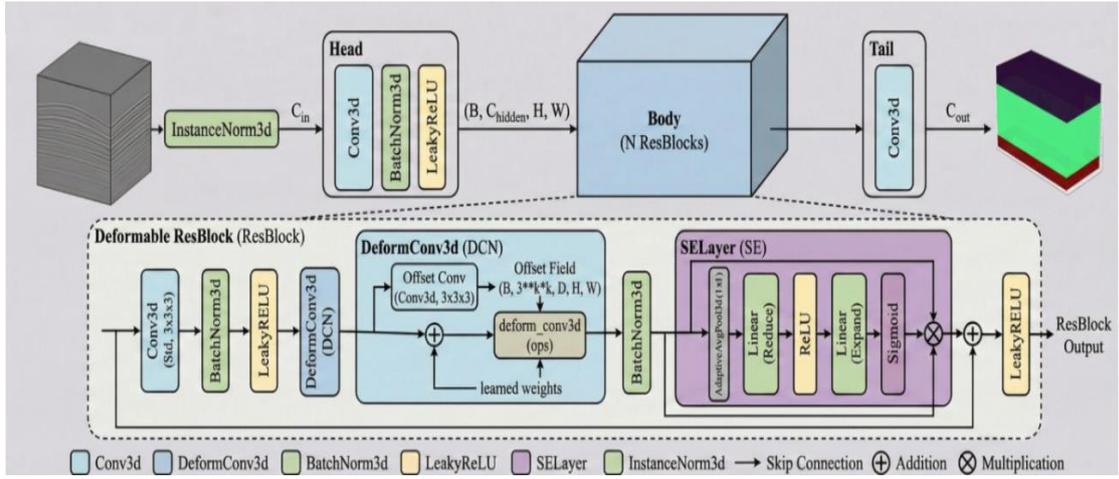

*Figure1 Network architecture diagram*

We apply the discrete cosine transform (DCT) to the seismic data. The sparse transform domain representations of the seismic data X and velocity model V can be defined as:

$$W_{seis} = \mathcal{R}(DCT(X)); W_{vel} = \mathcal{R}(DCT(V))$$

where $DCT(\cdot)$ represents the discrete cosine transform operator,; $W_{seis}$ and $W_{vel}$ represent the sparse transform domain coefficients of the seismic data and velocity model, respectively. $\mathcal{R}(\cdot)$ is the truncation operator, defined as:

$$\mathcal{R}(c_k) = \begin{cases} c_k, & \text{if } k \leq N \\ 0, & \text{if } k > N \end{cases}$$

In this formula, $c_k$ represents the coefficient of the k-th frequency component, and N is the preset truncation threshold. Figure 2 and Figure 3 present the results of applying DCT and IDCT to seismic data and the velocity model, as well as the corresponding residuals.

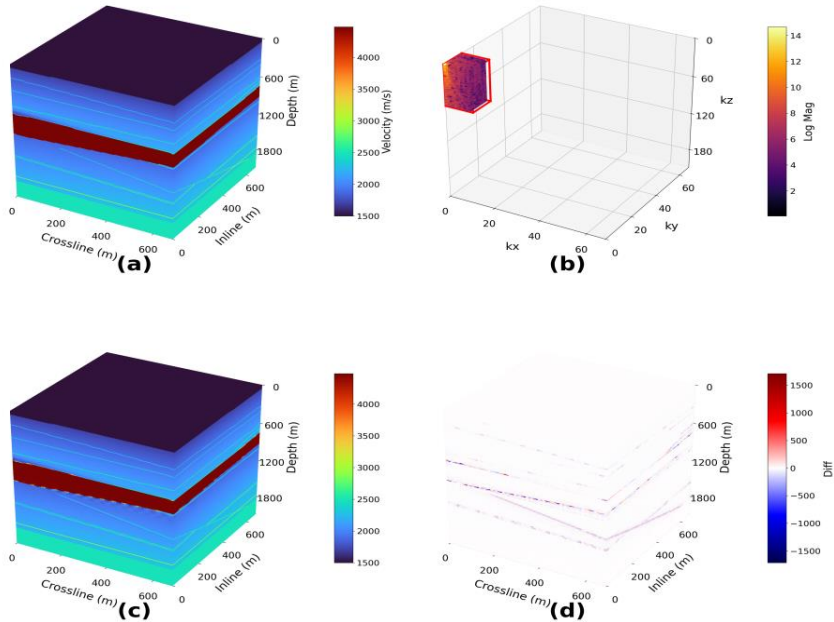

*Figure2 (a)Original velocity model; (b)Results after DCT; (c)Result of IDCT; (d)Residual error*

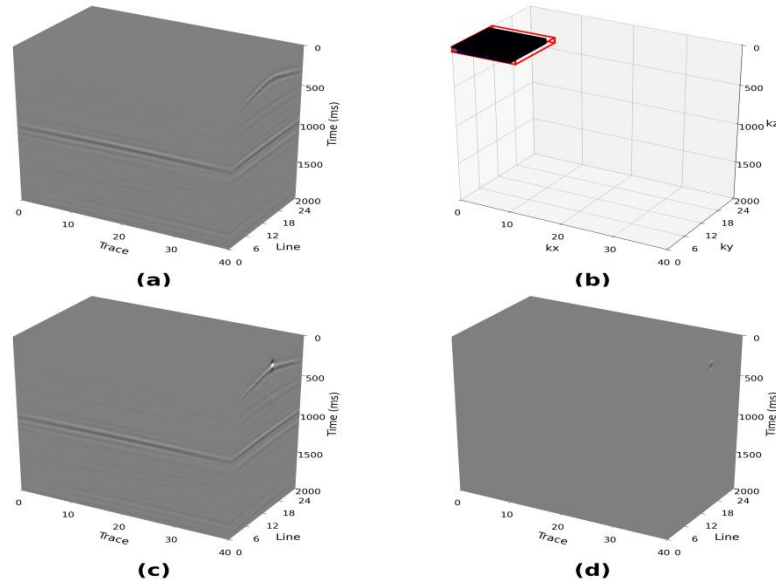

*Figure3 (a)Original seismic record; (b)Results after DCT; (c)Result of IDCT; (d)Residual error*

To validate the effectiveness of this method, we designed a controlled experiment: The control group used original seismic data and velocity models directly as input-label pairs for the neural network, while the experimental group utilized truncated DCT transform-domain coefficients of seismic data and velocity models as network inputs. This study obtained 676 3D velocity models by cropping the open 3D SEG/EAGE salt dome velocity model with complex salt dome structure (https://www.geoazur.fr/WIND/bin/view/Main/Data/WebHome), and then used finite difference to obtain the corresponding seismic data, which was then used for network training. For clear distinction and comparison, the method using DCT combined with the proposed network is denoted as DCT-SEDCN, and the method using only the network without DCT is denoted as SEDCN. All network structures, hyper-parameters, and training strategies are kept completely consistent between the two schemes, ensuring that the comparison only reflects the influence of DCT-based sparse domain transformation on inversion performance and computational efficiency. Table 1 shows the differences in inversion results and computational efficiency between the two methods.

*Table1 Comparison of Training and Testing Results between DCT-SEDCN and SEDCN Examples*

| Method | PSNR | SSIM | Total duration |
| --- | --- | --- | --- |
| DCT-SEDCN | 22.32 | 0.7085 | 14m28s |
| SEDCN | 19.56 | 0.7023 | 4h49m20s |

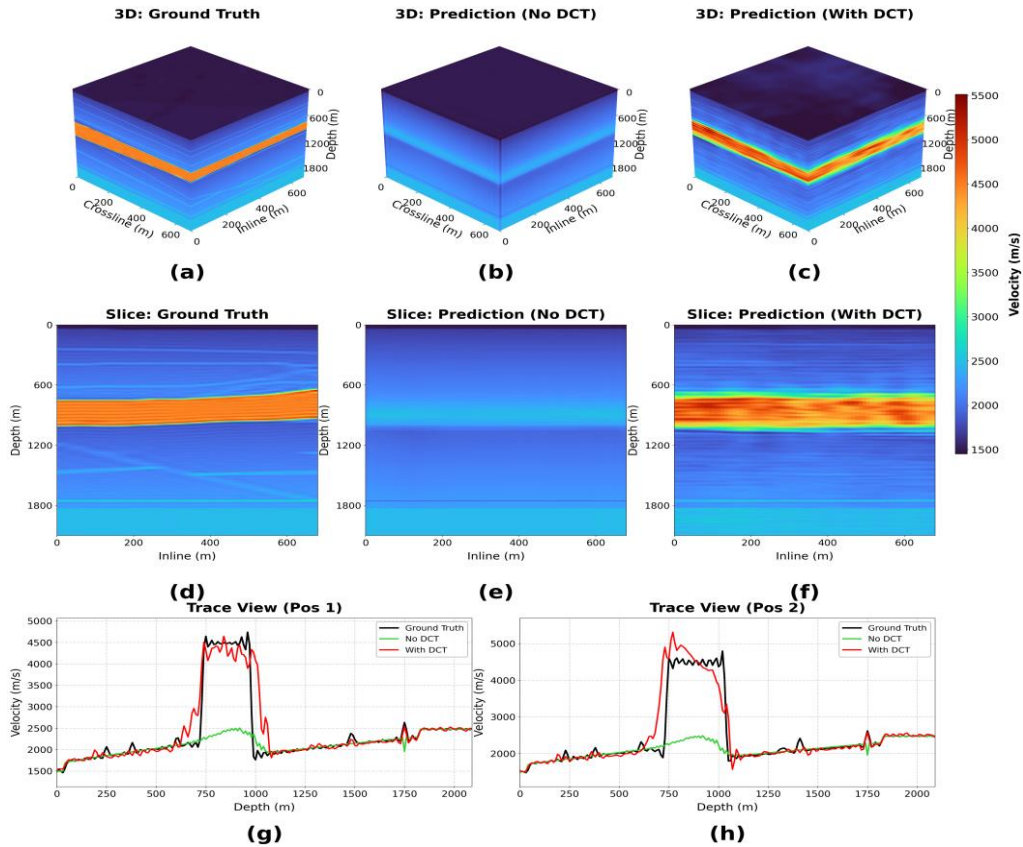

*Figure4 The seismic data inversion results are compared in three-dimensional visualization, YZ cross-section and single channel waveform ( DCT-SEDCN and SEDCN）.*

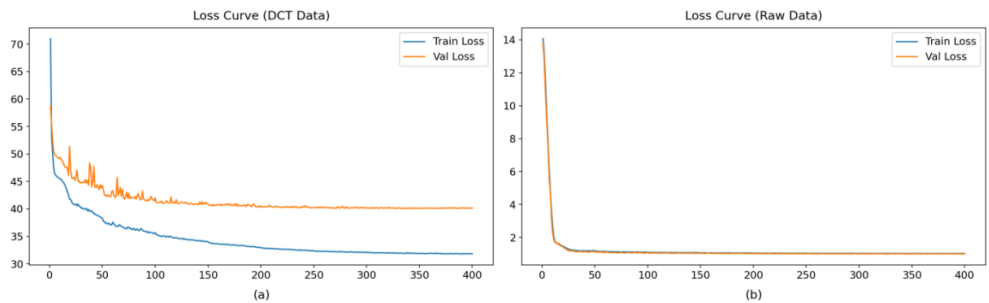

*Figure5 (a) DCT-SEDCN DCT Loss Curve; (b) SEDCN Loss Curve.*

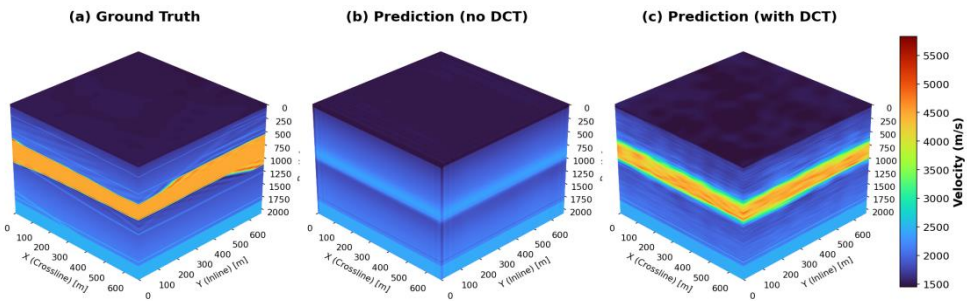

*Figure6 The other test dataset: (a) the true velocity model; (b) the spatiotemporal domain deep learning prediction results; (c) the DCT domain prediction results.*

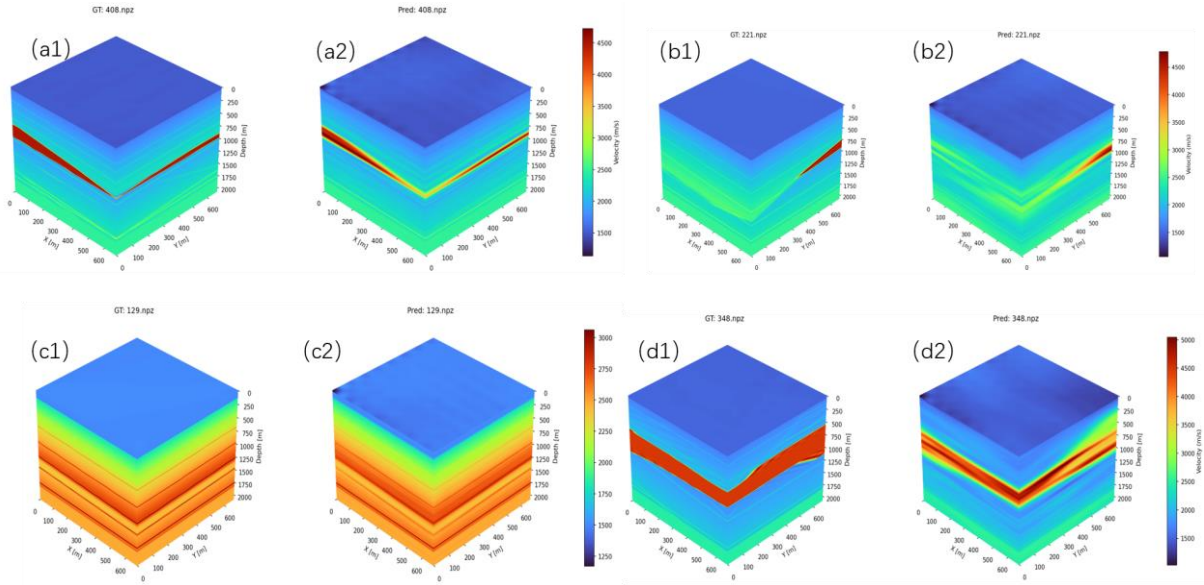

*Figure7 The other test dataset: (a1),(b1),(c1),(d1) the true velocity model; (a2),(b2),(c2),(d2) the DCT domain prediction results.*

## Results

As shown in Table 1, the proposed DCT deep learning method significantly improves both training efficiency and inversion accuracy compared with conventional deep learning algorithms in the spatiotemporal domain. The quantitative comparison indicates that the DCT-based approach not only achieves drastically higher computational efficiency but also provides more reliable and high-fidelity velocity model reconstruction.As illustrated in Figure 4, three-dimensional visualization, YZ cross-section slices, and single-trace waveform comparisons are comprehensively displayed to evaluate the reconstruction performance. The results obtained by conventional spatiotemporal-domain deep learning exhibit severe over-smoothing, blurred salt body boundaries, and indistinct velocity interfaces. The predicted velocity distribution fails to reflect the real structural characteristics, and obvious deviations can be observed in the single-trace waveforms. In contrast, the prediction results with DCT compression clearly recover the geometric shape and internal structure of the salt dome, with sharp boundaries and reasonable velocity gradients. The single-trace waveforms are highly consistent with the ground truth, demonstrating superior reconstruction fidelity and accuracy.

Figure 5 presents the loss curves of the two training strategies during the iteration process. The DCT-based method converges much faster and stabilizes at a significantly lower loss value, showing strong training stability. In comparison, the non-DCT method converges slowly and retains a higher final error, indicating that the network is difficult to capture effective structural features from high-dimensional raw data. This confirms that DCT sparse representation effectively accelerates model convergence and improves the reliability of model training.

As demonstrated in Figure 6, another independent test dataset is adopted for further validation. The prediction from the spatiotemporal domain method still fails to reflect fine structural details and shows excessive averaging, leading to inaccurate velocity estimation. In comparison, the DCT domain result accurately reconstructs the velocity distribution, preserves clear structural boundaries, and achieves stronger consistency with the true velocity model, verifying the strong generalization ability of the proposed method. Figure 7 further provides multiple groups of comparison slices from additional test samples. The true velocity models in panels (a1), (b1), (c1), and (d1) contain complex structural variations and strong velocity contrasts. The corresponding DCT-based predictions in panels (a2), (b2), (c2), and (d2) faithfully reproduce the key geological features, including structural boundaries, texture details, and velocity gradients, without obvious over-smoothing or distortion.

In summary, conventional spatiotemporal-domain deep learning struggles to capture effective high-frequency structural variations because of the high dimensionality of raw data, forcing the network to predict local average velocities to minimize statistical errors. In contrast, DCT compression filters out high-frequency noise and redundant information while retaining dominant low-frequency components that control major geological structures. This enables the network to focus on meaningful structural coefficients, leading to more accurate velocity prediction and higher geological realism.

## Discussion and Conclusions

Although the proposed DCT-based deep learning framework delivers considerable improvements in computational efficiency and inversion accuracy for 3D acoustic seismic velocity modeling, several inherent limitations exist and deserve thorough discussion. The current method is developed under the acoustic assumption, ignoring elastic properties, shear-wave splitting, and viscoelastic attenuation that are widely present in real earth media. As a result, it cannot recover multi-parameter information including P-wave velocity, S-wave velocity, density, and quality factor Q, which restricts its capability for quantitative reservoir interpretation. Meanwhile, the fixed truncation threshold used in DCT compression may lose weak high-frequency components related to small-scale geologic structures, potentially reducing spatial resolution in regions with strong velocity variations. In addition, the framework is only validated on synthetic SEG/EAGE 3D salt model data; its stability and generalization when facing field data issues such as strong noise, irregular acquisition geometry, and rugged topography still need systematic verification. To overcome these limitations and advance toward practical applications, we plan to extend the proposed method to 3D elastic and viscoelastic seismic multi-parameter inversion in future work. The core idea is to construct a physics-constrained multi-parameter deep learning framework under dual-domain sparse representation. We will replace the acoustic wave equation with elastic and viscoelastic wave equations in forward modeling to generate realistic multi-parameter training datasets that support P- and S-wave mode conversion, energy attenuation, and velocity dispersion. A unified

DCT sparse representation will be designed for multi-parameter models including P-wave velocity, S-wave velocity, density, and attenuation parameters, so as to compress redundant information and enhance the network's ability to capture structural correlations among different parameters. We will further introduce elastic wave equation constraints and petrophysical priors into the loss function to reduce cross-talk and non-uniqueness in multi-parameter inversion, ensuring that the outputs satisfy physical consistency.

The main challenges in extending to elastic and viscoelastic media include drastically increased computational cost, strong parameter coupling, non-unique solutions, and difficulties in synthetic-to-field generalization. To mitigate these issues, we will adopt DCT compression to reduce data dimensionality and accelerate both forward simulation and network training. We will introduce adaptive sparse truncation and parameter-specific regularization in the transform domain to alleviate trade-offs between elastic parameters. Semi-supervised learning and domain adaptation strategies will be integrated to reduce dependence on labeled synthetic data and improve performance on field data. We will also use mixed-precision computation and multi-scale inversion strategies to stabilize network training and improve convergence.

This study presents a dual-domain sparse representation deep learning framework based on discrete cosine transform (DCT) for efficient 3D seismic acoustic velocity inversion. By compressing raw seismic data and velocity models into a sparse transform domain, the method eliminates data redundancy, reduces input dimensionality, and shortens training time by more than 90% while significantly accelerating convergence. The DCT operation guides the network to focus on dominant structural frequency components rather than noisy temporal samples, thus extracting high-fidelity geological features more effectively. Compared with conventional deep learning inversion in the spatiotemporal domain, the proposed approach significantly mitigates the over-smoothing or "regression to the mean" effect, producing velocity models with clearer salt dome boundaries and finer textural details. These results confirm that DCT-based sparse representation combined with geometry-adaptive deep learning can effectively improve the efficiency, precision, and geological reliability of 3D seismic velocity inversion. The proposed framework provides a scalable and promising solution for large-scale 3D full-waveform inversion and will be further extended to elastic and viscoelastic multi-parameter inversion and field data applications in future research.

## Acknowledgements

This work was financially supported by the National Key R&D Program of China (2023YFF0803404), Zhejiang Provincial Natural Science Foundation (grant LY23D040001), Open Research Fund of Key Laboratory of Engineering Geophysical Prospecting and Detection of Chinese Geophysical Society (CJ2021GB01), Open Re-search Fund of Changjiang River Scientific Research Institute (CKWV20221011/KY), Zhou-Shan Science and Technology Project(grant 2023C81010), National Natural Science Foundation of China (grant 41904100).